\documentclass[12pt]{article}
\usepackage{amssymb,amsmath,epsfig,cite}

\normalbaselines

\begin{document}

\def\ep{\epsilon}
\def\p{\partial}

\title{New supersymmetric partners for the \\ associated Lam\'e 
potentials}

\author{David J. Fern\'andez C.${}^\dagger$ and Asish 
Ganguly${}^\ddagger$ \\ [5pt]
\small ${}^\dagger$Departamento de F\'{\i}sica, CINVESTAV \\
\small AP 14-740, 07000 M\'exico DF, Mexico \\ [5pt]
\small ${}^\ddagger$Department of Applied Mathematics, University of 
Calcutta \\
\small 92 Acharya Prafulla Chandra Road, Kolkata 700 009, India}

\date{}

\maketitle

\begin{abstract}
We obtain exact solutions of the one-dimensional Schr\"odinger
equation for some families of associated Lam\'e potentials with
arbitrary energy through a suitable ansatz, which may be
appropriately extended for other such a families. The formalism of
supersymmetric quantum mechanics is used to generate new exactly
solvable potentials.
\end{abstract}



\section{Introduction}

The solution of the one-dimensional Schr\"odinger equation with
periodic potentials is important due to the possibility of finding
interesting models which could be used in physics. This kind of
potentials admits a spectrum composed of allowed energy bands, in
which the physical eigenfunctions are bounded, separated by energy
gaps, where the eigenfunctions are unbounded and thus they cannot
have physical meaning due to their exponential growing when we
move far away from a given position.

Of special interest are the so-called exactly solvable problems
which, unfortunately, in the periodic case include very few
potentials, e.g., Lam\'e and some others. To be precise, by exact
solvability we mean that for the corresponding potential it is
possible to determine analytically the physical eigenfunctions for
energies in the allowed bands (the edges included), as well as the
non-physical solutions for energies in the gaps. It is important
to notice that in the direct spectral problems the unphysical
solutions are neglected because they are apparently useless.
However, in the inverse spectral problems, which somehow include
the supersymmetric quantum mechanics as a particular case
\cite{ba00,afhnns04}, they can be used as seeds to generate
exactly solvable potentials from a given initial one
\cite{mr04,nnr04,ac04,io04,su05,ff05}.

In recent times it has been realized that the associated Lam\'e
potentials, which include in particular the Lam\'e case, admit
explicit expressions for the band edge eigenfunctions
\cite{ks99,ga00,ks01,ga02,ks04}. Thus, it is natural to explore if
those potentials are exactly solvable in the sense pointed above.
If the answer turns out to be positive, then the band edge
eigenfunctions as well as the non-physical solutions for the gaps
can be used through supersymmetric quantum mechanics to generate
new exactly solvable potentials, some of which could provide some
interesting physical information.

In the next section we will consider some particular families of
the associated Lam\'e potentials and show that they belong to the
exactly solvable class. Then, the formalism of supersymmetric
quantum mechanics will be applied to generate new exactly solvable
potentials. We will end the paper with our conclusions.

\section{Associated Lam\'e potentials: general solutions}

The Schr\"odinger equation for the associated Lam\'e potentials in
Jacobi form may be expressed as
\begin{eqnarray}
&& -\frac{d^2\psi(x)}{dx^2} + V(x)\psi(x)=E\psi(x),
\qquad x\in \mathbb{R} ,\nonumber \\
&& V(x)=m(m+1)k^{2}{\rm sn}^{2}x + \ell(\ell+1)k^{2}\frac{{\rm
cn}^{2}x}{{\rm dn}^{2}x}. \label{asl}
\end{eqnarray}
Our aim is to find exact solutions of Eq.~(\ref{asl}) for
arbitrary values of $E$. Here ${\rm sn}x\equiv {\rm sn}(x,k), \
{\rm cn}x\equiv {\rm cn}(x,k), \ {\rm dn}x\equiv {\rm dn}(x,k)$
are three Jacobian elliptic functions of real modulus $k^2\
(0<k^{2}<1)$ and of double periods $4K,2iK';4K,4iK';$ $2K, 4iK'$
respectively, $K= \int_{0}^{\pi/2}
d\phi/\sqrt{1-k^{2}\sin^{2}\phi}$ is called complete elliptic
integral of second kind and $K'\:
=\int_{0}^{\pi/2}d\phi/\sqrt{1-k'^{2}\sin^{2}\phi}$,
$k'^{2}=1-k^{2}$ is known as complementary modulus. In the first
step of the process we express Eq.~(\ref{asl}) in Weierstrass form
to easily observe the difference from the Lam\'{e} potential:
\begin{equation}
-\frac{d^{2}\psi}{dz^{2}}+\left [m(m+1)\wp (z)
+\frac{\ell(\ell+1)\bar{e}_{2}\bar{e}_{3}}{\wp (z)-e_{1}}\right]
\psi =\tilde{E}\psi , \ \bar{e}_{i}=e_{1}-e_{i},\ i=2,3,
\label{aslw}
\end{equation}
where $z=(x-iK')/\sqrt{\bar{e}_{3}}\in \mathbb{C}, \ \psi(x)\equiv
\psi(z(x))$ and
$\tilde{E}=[E-\ell(\ell+1)]\bar{e}_{3}+e_{3}m(m+1)$. In
Eq.~(\ref{aslw}), $\wp(z)\equiv \wp(z;\omega,\omega')$ is the
Weierstrass elliptic function of half-periods
$\omega=K/\sqrt{\bar{e}_{3}}, \, \omega'=iK'/\sqrt{\bar{e}_{3}}$
and $\wp(\omega_{i})=e_{i}, \omega_{1}=\omega,
\omega_{2}=\omega+\omega', \omega_{3}=\omega'$. It is clear now
that the differential equation (\ref{aslw}) has regular
singularities at the poles of $\wp(z)$ and at points congruent to
$z=\omega_{1}$, the latter being absent in Lam\'{e} equation which
corresponds to $\ell =0$. Our aim is to find two linearly
independent solutions $\psi_{i}(z), \, i=1,2$ of Eq.~(\ref{aslw}).
It is well-known that $\Psi(z)=\psi_{1}(z)\psi_{2}(z)$ satisfies a
third order differential equation (see e.g. \cite{ww63})
\begin{eqnarray}
&& \frac{d^{3}\Psi}{dz^{3}}-4\left [m(m+1)\wp(z)
+\frac{\ell(\ell+1)\bar{e}_{2}\bar{e}_{3}}{\wp(z)-e_{1}}-\tilde{E}\right]
\frac{d\Psi}{dz} 
\nonumber \\
&& \mbox{}\hspace{1cm} 
- 2\left [m(m+1)-\frac{\ell(\ell+1)
\bar{e}_{2}\bar{e}_{3}}{\{\wp(z)-e_{1}\}^{2}}\right]\wp'(z)\Psi=0
, \label{prod}
\end{eqnarray}
where throughout this article the prime (except in $K'$ and
$\omega'$) will denote derivative with respect to the shown
argument. To obtain its solutions we propose an ansatz and use a
fitting procedure to automatically fix $m$, $\ell$ and also the
unknown quantities, namely;
\begin{equation}
\Psi(z)=[\wp(z)-e_{1}]+A_{1} + \frac{A_{2}}{\wp(z)-e_{1}}.
\label{ansz1}
\end{equation}
Let us notice that this ansatz is motivated by realizing that
Eq.~(\ref{prod}) has an additional regular singularity at
$z=\omega_{1}$ compared to Lam\'{e} case ($\ell=0$), and so it is
expected that its solution will also have a singularity at this
point. Substituting (\ref{ansz1}) into (\ref{prod}) and equating
coefficients of each power of $[\wp(z)-e_{1}]$ to zero, we find
the following solutions for $A_{1},A_{2},m$ and $\ell$:
\begin{eqnarray}
a) \quad m=1,-2; \ \ell=1,-2; &
\hspace{.5cm}A_{1}=\tilde{E}+e_{1},
                & \hspace{.5cm}A_{2}=\bar{e}_{2}\bar{e}_{3} \, ,
\label{sol1,1} \\
b) \quad m=1,-2; \ \ell=0,-1; &
\hspace{.5cm}A_{1}=\tilde{E}+e_{1},
                    & \hspace{.5cm}A_{2}=0. \label{sol1,2}
\end{eqnarray}
The solutions (\ref{sol1,1}) and (\ref{sol1,2}) correspond to
eight points in the $m-\ell$ plane, four of which (for $\ell=0, \,
-1$) lead to a Lam\'{e} potential. Of the other four points, the
one $(1,1)$ lies in our effective region ($m,l\geq -1/2, \ m\geq
\ell$). Thus, the ansatz (\ref{ansz1}) will give us the general
solution of the associated Lam\'{e} potential with $m=\ell=1$ for
arbitrary $\tilde{E}$.

The ansatz (\ref{ansz1}) may be extended by inserting additional
terms involving higher positive and negative powers of
$[\wp(z)-e_{1}]$. For instance, one may take
\begin{equation}
\Psi(z)=[\wp(z)-e_{1}]^{2}+B_{1}[\wp(z)-e_{1}]+B_{2} +
\frac{B_{3}}{\wp(z)-e_{1}} \, , \label{ansz2}
\end{equation}
for which the solutions become
\begin{eqnarray}
\hskip-1.5cm && c) \: m = 2,-3; \ \ell  =  1,-2; \quad B_{1} 
=  2e_{1}  +  \frac{\tilde{E}}3, \quad
B_{2} = \left(\frac{\tilde{E}}3  -  e_{1}\right) B_{1}, 
\qquad \qquad  B_{3} =  \frac{\bar{e}_{2}\bar{e}_{3}B_{1}}3,  
\label{sol2,1} \\
\hskip-1.3cm && d) \: m  =  2,-3; \ \ell  =  0,-1; \quad 
B_{1} =  2e_{1}  +  \frac{\tilde{E}}3, \quad
B_{2} = \left(\frac{\tilde{E}}3  -  e_{1}\right) B_{1}
+ \bar{e}_{2}\bar{e}_{3}, \quad B_{3}  =  0 . \label{sol2,2}
\end{eqnarray}
Once again, out of the eight points four correspond to a Lam\'{e}
model and in our effective region we have the point $(2,1)$.

We have found that the product of solutions of the associated
Lam\'{e} equation (\ref{aslw}) takes the form:
\begin{equation}
\Psi(z)=\frac{1}{\wp(z)-e_{1}}\prod_{r=1}^{2}[\wp(z)-\wp(a_{r})],
\quad m=\ell=1 \, , \label{prod1,1}
\end{equation}
\begin{equation}
\Psi(z)=\frac{1}{\wp(z)-e_{1}}\prod_{r=1}^{3}[\wp(z)-\wp(b_{r})],
\quad m=2,\ \ell=1 \, . \label{prod2,1}
\end{equation}
In Eqs.~(\ref{prod1,1}),(\ref{prod2,1}) the quantities
$\wp(a_{r}), \, r=1,2$ and $\wp(b_{r}), \, r=1,2,3$ are the zeros
of the polynomials arising in the numerators of (\ref{ansz1}) and
(\ref{ansz2}), and this implies that the determination of $a_{r}$
and $b_{r}$ involves transcendental equations of the type
$\wp(t)=c$. Since $\wp(z)$ is an even function, the resulting
ambiguity of signs in $a_{r},b_{r}$ has to be fixed from the
convention $\Psi'(z)>0$ at the points $z=a_{r},b_{r}$.

Once we know the product of solutions, it is straightforward to
obtain two linearly independent solutions for the associated
Lam\'{e} equation (\ref{asl}) following the same procedure adopted
for Lam\'{e} \cite{ww63}. Up to some constant factors, our final
results are as follows:
\begin{equation}
\hspace{-8.5cm}\textbf{1.}\: V(x)=2k^{2}{\rm
sn}^{2}x+2k^{2}\frac{{\rm cn}^{2}x}{{\rm dn}^{2}x} \label{asl1,1}
\end{equation}
\vskip-0.5cm
\begin{eqnarray}
\psi_{1,2}(x) & = & \frac{\prod^{2}_{r=1}
\sigma\left(\frac{x-iK'}{\sqrt{\bar{e}_{3}}} \pm a_{r} \right)}{
\sigma\left(\frac{x-iK'}{\sqrt{\bar{e}_{3}}}+ \omega_1\right)
\sigma\left(\frac{x-iK'}{\sqrt{\bar{e}_{3}}}\right)}
\exp\left[\mp\frac{x}{\sqrt{\bar{e}_{3}}} \sum^{2}_{r=0}\zeta
(a_{r})\right] \, . \label{gs1,1,1}
\end{eqnarray}
\begin{equation}
\hspace{-8.5cm}\textbf{2.}\: V(x)=6k^{2}{\rm
sn}^{2}x+2k^{2}\frac{{\rm cn}^{2}x}{{\rm dn}^{2}x} \label{asl2,1}
\end{equation}
\vskip-0.5cm
\begin{eqnarray}
\psi_{1,2}(x) & = & \frac{\prod^{3}_{r=1}
\sigma\left(\frac{x-iK'}{\sqrt{\bar{e}_{3}}} \pm b_{r} \right)}{
\sigma\left(\frac{x-iK'}{\sqrt{\bar{e}_{3}}}+ \omega_1\right)
\sigma^2\left(\frac{x-iK'}{\sqrt{\bar{e}_{3}}}\right)} \exp
\left[\mp\frac{x}{\sqrt{\bar{e}_{3}}} \sum^{3}_{r=0}\zeta
(b_{r})\right] \, . \label{gs2,1,1}
\end{eqnarray}
In above solutions we have taken $a_0=b_0\equiv \mp \omega_1$,
$\sigma(x)$ and $\zeta(x)$ are the quasi-periodic Weierstrass
elliptic sigma and zeta functions, which are defined by
$\zeta'(x)=-\wp(x)$, $[\ln \sigma(x)]'=\zeta (x)$. It is not very
difficult to check that both solutions (\ref{gs1,1,1}) become
proportional to the three band edge eigenfunctions for the
associated Lam\'e potentials with $m=\ell = 1$ when $E$ takes the
three eigenvalues $E_{0,1} = 2 + k^2 \mp 2 k', \ E_2=4$, which is
consistent with the fact that this potential has one finite energy
band and one finite gap \cite{ga00,ga02}. The realization that the
associated Lam\'e potential for $m=\ell=1$ and modulus parameter
$k^2$ is exactly solvable was recently found by noticing
\cite{ks04} that it can be obtained from the Lam\'e one [take
$m=1, \, \ell=0$ and a transformed modulus parameter in
Eq.~(\ref{asl})] via a well-known Landen transformation
\cite{ww63}. In contrast, here we have proved that property by
directly finding the general solution for arbitrary $E\in{\mathbb
R}$. On the other hand, up to constant factors both solutions
(\ref{gs2,1,1}) reduce to the five band edge eigenfunctions for
the associated Lam\'e potentials with $m=2, \ \ell=1$ when $E$
takes the five eigenvalues $E_0=4k^2, \ E_{1,4} = 5 + k^2 \mp 2
\sqrt{4-3k^2}, \ E_{2,3} = 5 + 2k^2 \mp 2 \sqrt{k^4-5 k^2 +4}$
\cite{ga00,ga02}. This is again consistent with the spectral
properties for this associated Lam\'e potential which has two
finite energy bands and two finite gaps. To the best of our
knowledge, the fact that the associated Lam\'e potentials for
$m=2,\ell=1$ are exactly solvable was previously unknown.

\section{Supersymmetric partner potentials}

In the modern approach to the first-order supersymmetric quantum
mechanics (SUSY QM), in which a first order differential
intertwining operator is used to implement the transformation, the
seed Schr\"odinger solution $u(x)$ can be either physical or
non-physical but it has to be nodeless to avoid singularities in
the new potential $\widetilde V(x) = V(x) - 2 [\ln u(x)]''$ (see
the collection of articles in \cite{afhnns04}). This is achieved
by taking the factorization energy $\epsilon$ such that
$\epsilon\leq E_0$, where $E_0$ is the lowest band edge energy. In
particular, if we chose $u(x)$ as any of the two Bloch functions
$\psi_{1,2}(x)$ derived in the previous section then $\widetilde
V(x)$ will be periodic, with the same band spectrum as $V(x)$. On
the other hand, if we choose $u(x)$ as a nodeless linear
combination of the two Bloch functions $\psi_{1,2}(x)$ associated
to $\epsilon$, then $\widetilde V(x)$ will present a periodicity
defect, and the spectrum of $\widetilde H$ will consist of the
allowed energy bands of $H$ plus an isolated bound state at
$\epsilon$, for which the corresponding eigenfunction $1/u(x)$
will be square-integrable \cite{tr89,fmrs02a,fmrs02b,ro03}.

Let us mention that for the second order SUSY QM, which involves
differential intertwining operators of second order, the key
function which has to be nodeless is the Wronskian of the two seed
solutions $u_1(x), \ u_2(x)$ associated to the factorization
energies $\epsilon_1, \ \epsilon_2$
\cite{ff05,fmrs02a,fmrs02b,ro03,bgbm99,fnn00}. In this frame it is
possible to use solutions for which $\epsilon_1, \ \epsilon_2$
have unexpected positions, e.g., both can lie in a finite energy
gap and produce however non-singular SUSY transformations. To be
brief, in this letter we will not apply the second order SUSY
transformations and we will restrict our discussion to the first
order cases mentioned above (see however \cite{fg05}).

By applying now the first order SUSY algorithm to the associated
Lam\'e potentials with $m=\ell=1$ and $m=2, \ \ell =1$, using as
seed solution any of the Bloch functions given by
Eqs.~(\ref{gs1,1,1}) and (\ref{gs2,1,1}) respectively, we arrive
to the following new exactly solvable periodic potentials which
are strictly isospectral to the corresponding associated Lam\'e
potential:

\begin{equation}
\hspace{-8.5cm}\textbf{1.}\: m=\ell=1 \nonumber
\end{equation}
\vskip-0.8cm
\begin{eqnarray}
\widetilde V(x) = && -4k^{2}{\rm sn}^{2}x + 2 \sum_{r=1}^{2}\left[
\frac{{\rm sn}x\,{\rm cn}x\,{\rm dn}x\pm
\beta(a_r\sqrt{\bar{e}_{3}})}{{\rm sn}^{2}x +
\alpha^2(a_r\sqrt{\bar{e}_{3}})}
\right] \nonumber \\
&& + 2\left\{k^2\left[2+
\sum_{r=1}^{2}\alpha^2(a_r\sqrt{\bar{e}_{3}})\right] + 2 \right\}
\label{nv1,1}
\end{eqnarray}
\begin{equation}
\hspace{-8.5cm}\textbf{2.}\: m=2, \ \ell=1 \nonumber
\end{equation}
\vskip-0.8cm
\begin{eqnarray}
\widetilde V(x) = && -4k^{2}{\rm sn}^{2}x + 2 \sum_{r=1}^{3}\left[
\frac{{\rm sn}x\,{\rm cn}x\,{\rm dn}x\pm
\beta(b_r\sqrt{\bar{e}_{3}})}{{\rm sn}^{2}x +
\alpha^2(b_r\sqrt{\bar{e}_{3}})}
\right] \nonumber \\
&& + 2\left\{k^2\left[3 +
\sum_{r=1}^{3}\alpha^2(b_r\sqrt{\bar{e}_{3}})\right] + 3 \right\}
\label{nv2,1}
\end{eqnarray}
where $\alpha^2(\tau) = -1/(k^{2} \, {\rm sn}^2\tau), \
\beta(\tau) = -{\rm cn}\,\tau\ {\rm dn}\,\tau/(k^{2}\,{\rm
sn}^{3}\tau)$.

\begin{figure}[ht]\centering
\epsfig{file=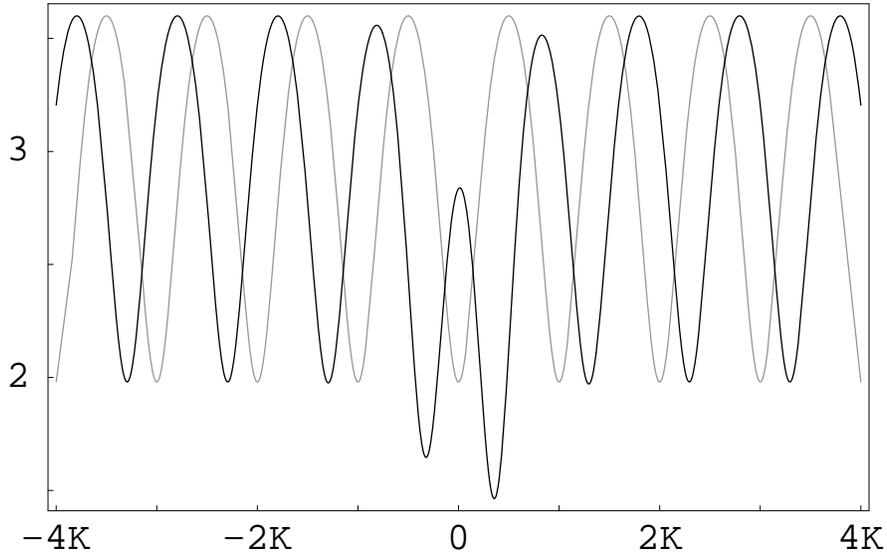, width=12cm} \caption{\small Potential
$\widetilde V(x)$ with a periodicity defect (black curve)
generated from the associated Lam\'e potential with $m=\ell=1$
(gray curve), $k^2=0.99$, and $u(x) = \psi_1(x) + 3 \psi_2(x)/2$
with $\epsilon = 2.4$, which leads to $a_1=-1.089, \ a_2=2.607$.
The new Hamiltonian $\widetilde H$ has an extra bound state at
$\epsilon$.}
\end{figure}

On the other hand, if the seed solution is a linear combination of
the two positive definite Bloch functions, $u(x) = \psi_1(x) +
\lambda \psi_2(x)$, then for $\lambda<0$, $u(x)$ will have always
a node, inducing then a singular SUSY transformation. For $\lambda
=0$ and $\lambda = \infty$ we will recover once again the
previously discussed case when $u(x)$ is one of the Bloch
functions. Finally, it is interesting to observe that for
$\lambda>0$, $u(x)$ will be nodeless, and the new potential will
not be strictly periodic (it will have a periodicity defect). The
spectrum for the SUSY generated potential $\widetilde V(x)$ will
consist of the allowed energy bands of the initial potential plus
an isolated level embedded in the infinite region below the
`ground state energy' of the initial potential. Unfortunately,
here the explicit expressions for $\widetilde V(x)$ are not
compact. Thus, we decided to illustrate this case graphically, and
an example of the SUSY partner potential with one periodicity
defect of the associated periodic Lam\'e potential for $m=\ell=1$
is given in Fig.~1 while for $m=2, \ \ell =1$ the corresponding
SUSY partner potentials are shown in Fig.~2.

\begin{figure}[ht]\centering
\epsfig{file=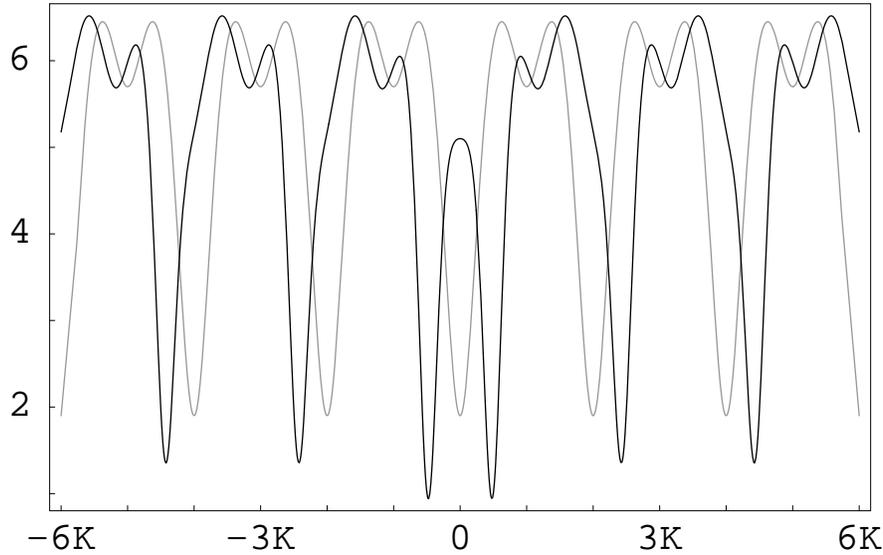, width=12cm} \caption{\small Potential
with a periodicity defect (black curve) produced from the
associated Lam\'e potential with $m=2, \ \ell=1$ (gray curve),
$k^2=0.95$, and $u(x) = \psi_1(x) + \psi_2(x)$ with $\epsilon =
3.5$, which gives $b_1=-2.392, \ b_2= 1.26 + 0.614 \, i, \
b_3=b_2^*$. The new Hamiltonian $\widetilde H$ has an extra bound
state at $\epsilon$.}
\end{figure}

\section{Conclusions}

Through an appropriate ansatz, we have solved the stationary
Schr\"odinger equation for the associated Lam\'e potentials with
an arbitrary energy corresponding to the parameter pairs $(1,1)$
and $(2,1)$. This suggests that the associated Lam\'e equation
with any integer values of the potential parameters is exactly
solvable. This assertion can be proved case by case by
appropriately modifying the ansatz, given by
Eqs.~(\ref{ansz1}),(\ref{ansz2}) and using a fitting procedure to
find the corresponding analytic solution. For instance, one may
fit the solution $\Psi (z)=\sum_{r=-2}^{3}C_{r}[\wp
(z)-e_{1}]^{r}$ to (\ref{prod}), which will effectively correspond
to the point (3,2) in the $m-\ell$ plane, and so on. However, one
of the aims of this paper was to show that this can be done
through the simplest non-trivial available examples. On the other
hand, the first order supersymmetry transformations were used to
generate new exactly solvable potentials which can be either
periodic or with a periodicity defect, depending on how we choose
the seed Schr\"odinger solutions. This kind of SUSY
transformations, together with the second order ones, represent
the most simple theoretical tools for designing potentials with
prescribed spectra, a subject which every day seems closer to
physical reality.

\section*{Acknowledgements}

The authors acknowledge the support of Conacyt, project No.
40888-F. One of us (AG) acknowledges the warm hospitality at
Departamento de F\'{\i}sica, Cinvestav, and also thanks City
College authorities for a study leave.


\begin{thebibliography}{99}

\bibitem{ba00} B.K. Bagchi, {\it Supersymmetry in Quantum and Classical
Mechanics}, Chapman and Hall, Boca Raton, Florida (2000).

\bibitem{afhnns04} I. Aref'eva, D.J. Fern\'andez, V. Hussin, J. Negro,
L.M. Nieto, B.F. Samsonov, {\it Special issue dedicated to the
subject of the International Conference on Progress in
supersymmetric quantum mechanics}, J. Phys. A {\bf 37}, Number 43
(2004).

\bibitem{mr04} B. Mielnik, O. Rosas-Ortiz, J. Phys. A {\bf 37} (2004)
10007.

\bibitem{nnr04} J. Negro, L.M. Nieto, O. Rosas-Ortiz, J. Phys. A {\bf 37}
(2004) 10079.

\bibitem{ac04} A.A. Andrianov, F. Cannata, J. Phys. A {\bf 37} (2004)
10297.

\bibitem{io04} M.V. Ioffe, J. Phys. A {\bf 37} (2004) 10363.

\bibitem{su05} C.V. Sukumar, AIP Conference Proceedings {\bf 744} (2005)
166.

\bibitem{ff05} D.J. Fern\'andez, N. Fern\'andez-Garc\'{\i}a, AIP
Conference Proceedings {\bf 744} (2005) 236 (quant-ph/0502098).

\bibitem{ks99} A. Khare, U. Sukhatme, J. Math. Phys. {\bf 40} (1999) 5473.

\bibitem{ga00} A. Ganguly, Mod. Phys. Lett. A {\bf 15}
(2000) 1923 (math-ph/0204026).

\bibitem{ks01} A. Khare, U. Sukhatme, J. Math. Phys. {\bf 42} (2001) 5652.

\bibitem{ga02} A. Ganguly, J. Math. Phys. {\bf 43} (2002) 1980
(math-ph/0207028); {\it ibid} {\bf 43} (2002) 5310
(math-ph/0212045).

\bibitem{ks04} A. Khare, U. Sukhatme, J. Phys. A {\bf 37} (2004) 10037.

\bibitem{ww63} E.T. Whittaker, G.N. Watson, {\it A course of modern 
analysis}, Cambridge University Press, Cambridge (1963).

\bibitem{tr89} L. Trlifaj, Inv. Prob. {\bf 5} (1989) 1145.

\bibitem{fmrs02a} D.J. Fern\'andez, B. Mielnik, O. Rosas-Ortiz, B.F.
Samsonov, J. Phys. A {\bf 35} (2002) 4279 (quant-ph/0303051).

\bibitem{fmrs02b} D.J. Fern\'andez, B. Mielnik, O. Rosas-Ortiz, B.F.
Samsonov, Phys. Lett. A {\bf 294} (2002) 168 (quant-ph/0302204).

\bibitem{ro03} O. Rosas-Ortiz, Rev. Mex. F\'{\i}s. {\bf 49 S2} (2003) 145
(quant-ph/0302189).

\bibitem{bgbm99} B. Bagchi, A. Ganguly, D. Bhaumik, A. Mitra, Mod. Phys.
Lett. A {\bf 14} (1999) 27.

\bibitem{fnn00} D.J. Fern\'andez, J. Negro, L.M. Nieto, Phys. Lett. A {\bf
275} (2000) 338.

\bibitem{fg05} D.J. Fern\'andez, A. Ganguly, in preparation.

\end{thebibliography}
\end{document}